\PassOptionsToPackage{unicode}{hyperref}
\PassOptionsToPackage{hyphens}{url}
\PassOptionsToPackage{dvipsnames,svgnames,x11names}{xcolor}
\documentclass[
  letterpaper,
  DIV=11,
  numbers=noendperiod]{scrartcl}

\usepackage{amsmath,amssymb}
\usepackage{iftex}
\ifPDFTeX
  \usepackage[T1]{fontenc}
  \usepackage[utf8]{inputenc}
  \usepackage{textcomp} 
\else 
  \usepackage{unicode-math}
  \defaultfontfeatures{Scale=MatchLowercase}
  \defaultfontfeatures[\rmfamily]{Ligatures=TeX,Scale=1}
\fi
\usepackage{lmodern}
\ifPDFTeX\else  
\fi
\IfFileExists{upquote.sty}{\usepackage{upquote}}{}
\IfFileExists{microtype.sty}{
  \usepackage[]{microtype}
  \UseMicrotypeSet[protrusion]{basicmath} 
}{}
\makeatletter
\@ifundefined{KOMAClassName}{
  \IfFileExists{parskip.sty}{%
    \usepackage{parskip}
  }{
    \setlength{\parindent}{0pt}
    \setlength{\parskip}{6pt plus 2pt minus 1pt}}
}{
  \KOMAoptions{parskip=half}}
\makeatother
\usepackage{xcolor}
\ifx\paragraph\undefined\else
  \let\oldparagraph\paragraph
  \renewcommand{\paragraph}[1]{\oldparagraph{#1}\mbox{}}
\fi
\ifx\subparagraph\undefined\else
  \let\oldsubparagraph\subparagraph
  \renewcommand{\subparagraph}[1]{\oldsubparagraph{#1}\mbox{}}
\fi

\usepackage{longtable,booktabs,array}
\usepackage{calc} 
\usepackage{etoolbox}
\makeatletter
\patchcmd\longtable{\par}{\if@noskipsec\mbox{}\fi\par}{}{}
\makeatother
\IfFileExists{footnotehyper.sty}{\usepackage{footnotehyper}}{\usepackage{footnote}}
\makesavenoteenv{longtable}
\usepackage{graphicx}
\makeatletter
\def\maxwidth{\ifdim\Gin@nat@width>\linewidth\linewidth\else\Gin@nat@width\fi}
\def\maxheight{\ifdim\Gin@nat@height>\textheight\textheight\else\Gin@nat@height\fi}
\makeatother
\setkeys{Gin}{width=\maxwidth,height=\maxheight,keepaspectratio}
\makeatletter
\def\fps@figure{htbp}
\makeatother

\KOMAoption{captions}{tableheading}
\makeatletter
\@ifpackageloaded{caption}{}{\usepackage{caption}}
\AtBeginDocument{%
\ifdefined\contentsname
  \renewcommand*\contentsname{Table of contents}
\else
  \newcommand\contentsname{Table of contents}
\fi
\ifdefined\listfigurename
  \renewcommand*\listfigurename{List of Figures}
\else
  \newcommand\listfigurename{List of Figures}
\fi
\ifdefined\listtablename
  \renewcommand*\listtablename{List of Tables}
\else
  \newcommand\listtablename{List of Tables}
\fi
\ifdefined\figurename
  \renewcommand*\figurename{Figure}
\else
  \newcommand\figurename{Figure}
\fi
\ifdefined\tablename
  \renewcommand*\tablename{Table}
\else
  \newcommand\tablename{Table}
\fi
}
\@ifpackageloaded{float}{}{\usepackage{float}}
\floatstyle{ruled}
\@ifundefined{c@chapter}{\newfloat{codelisting}{h}{lop}}{\newfloat{codelisting}{h}{lop}[chapter]}
\floatname{codelisting}{Listing}

\makeatother
\makeatletter
\@ifpackageloaded{caption}{}{\usepackage{caption}}
\@ifpackageloaded{subcaption}{}{\usepackage{subcaption}}
\makeatother
\ifLuaTeX
  \usepackage{selnolig}  
\fi
\usepackage{bookmark}

\IfFileExists{xurl.sty}{\usepackage{xurl}}{} 
\hypersetup{
  pdftitle={Minute-by-Minute: Financial Markets' Reaction to the 2020 U.S. Election},
  pdfauthor={Matthew DeHaven; Hannah Firestone; Chris Webster},
  colorlinks=true,
  linkcolor={blue},
  filecolor={Maroon},
  citecolor={Blue},
  urlcolor={Blue},
  pdfcreator={LaTeX via pandoc}}

\title{Minute-by-Minute: Financial Markets' Reaction to the 2020 U.S.
Election}
\author{Matthew DeHaven \and Hannah Firestone \and Chris Webster}
\date{2024-07-03}

\begin{document}
\maketitle

Authors' note: This short article was written June 2021.

\section[Introduction]{\texorpdfstring{Introduction\footnote{We thank
  Ricardo Correa and Juan M. Londono for their useful comments. All
  errors remain our own.}}{Introduction}}\label{introduction}

The 2020 U.S. presidential election garnered notable global attention
from market participants, which was reflected, to some degree, in
immediate changes in asset prices. Since financial markets respond
rapidly to new and diverse information, it is difficult to disentangle
any one event's impact. However, the election offers a unique
opportunity to overcome this identification obstacle: major results came
in overnight, a sufficiently small window that concurrent events were
limited in scope. Using a novel data source to get minutely estimates of
probabilities for the presidential outcome, we examine the relationship
between the 2020 election and various financial markets.

We find striking correlations between the presidential election outcome
probability and major financial indicators, including USD currency
pairs, bond prices, stock index futures, and a market volatility
measure. The correlations are consistent with `risk-on' behavior in
markets, a term which describes investors moving toward riskier asset
classes, as the election results became clearer. Further, we decompose
the market reaction into a `reduction in uncertainty' component and a
`probability of a Democratic party presidency' component. This
decomposition reveals how markets reacted to the increasing certainty of
the outcome as election results came in. Finally, we analyze the
differing market reactions to the presidential election and the Senate
election, including data from the unique Georgia runoffs, and
demonstrate that bond prices were particularly sensitive to the
probability of a combined Democratic Senate and Presidency.

\section{Background and Data}\label{background-and-data}

Prediction markets offer a unique source of data on political and
economic events. Betting on U.S. elections has been commonplace since at
least the late 1800s (Rhode \& Strumpf, 2004; Wolfers, 2009), and the
advent of the internet in the 1990s introduced centralized `idea
futures' markets with many users (Hanson, 1996). These markets match two
users to purchase a `yes' and `no' share for a contract (that is, ``Will
candidate X be elected?'') at a price ranging from 0 to 1. In an
efficient market, the resulting price of a `yes' share can be
interpreted as the probability of the event `X' happening. For example,
if the price of a `yes' share was \$0.60 and a user predicted the true
probability of the event to be 0.9, the user would buy a `yes' share,
making an expected profit of \$0.30 per share. If many buyers expected
the same probability of 0.9 the price of the `yes' share would rise
closer to \$0.90. With enough users and enough information, the market
price converges to the expected probability of the event occurring.

Since the popularization of these betting markets in the 1990s, there
has been considerable research on their utility in predicting economic
variables (Hahn \& Tetlock, 2005; Snowberg et al., 2013), forecasting
elections (Arrow et al., 2008; Reade \& Vaughan Williams, 2019), and
aiding corporate decision making (O'Leary, 2015; Plott \& Chen, 2002;
Spears et al., 2009). The data in these studies is often at the daily
level, or even more infrequent. In this note, we take advantage of a
novel source of minutely prediction market data from PredictIt.
PredictIt is an online prediction market in the U.S., offering markets
on various U.S. and global political events. For each election
(i.e.~party control of the Senate and Presidency), we get minutely data
on the market `yes' and `no' prices, and we take the most recent trade
to arrive at the market-implied event probability.

To isolate financial markets' reaction to the election, we narrow the
time range to election night (November 3, 2020) and the following day.
Over the course of the night, there was a remarkable shift in the
perceived probability of either the Democratic or Republican candidate
winning the presidential election. The Democratic candidate's
probability of winning the election (Democratic Presidency Probability)
ranged from about 20\% to 90\% over the course of 30 hours, providing
enough variation to find statistically significant relationships. As
shown in Panel A of Figure 1, at 6:00 p.m. on November 3, the Democratic
candidate's probability of winning the presidency was close to 70\%. As
results were published, the probability fell all the way to 20\%, before
steadily rebounding late into the night.

\begin{figure}

\centering{

\includegraphics[width=5in,height=\textheight]{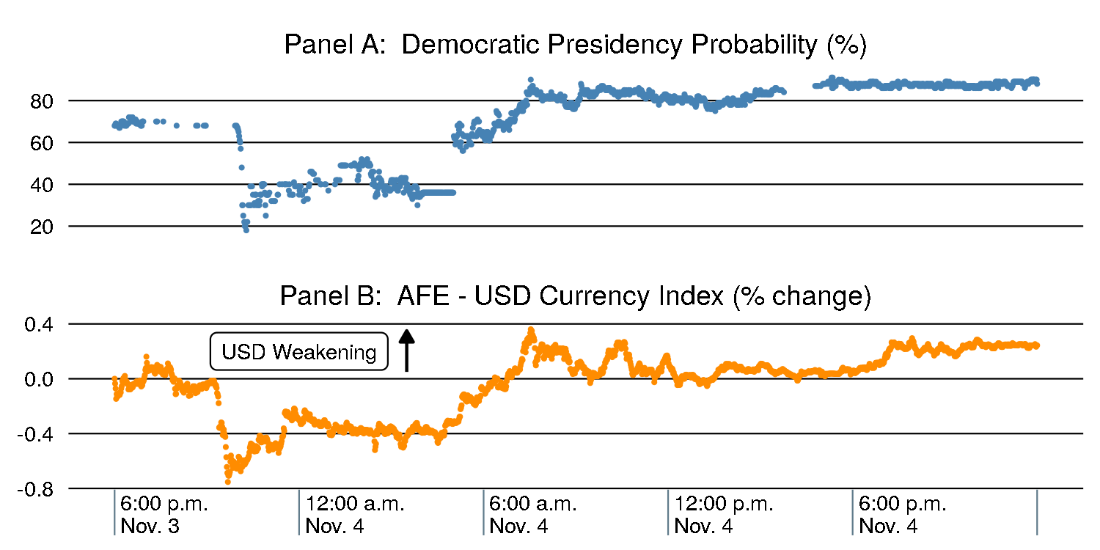}

Note: Observations are by the minute. The Democratic Presidency
Probability shows the latest traded price (which can be interpreted as a
probability) from PredictIt. The AFE-USD currency index is represented
as a percent change from the value before 6:00 p.m. on November 3rd,
2020, with higher values denoting the depreciation of the dollar.

Source: PredictIt, Bloomberg.

}

\caption{\label{fig-1}Markets during Election Night 2020.}

\end{figure}%

These large shifts reflect two factors: (1) the variable times at which
states reported results, and (2) the realization of a large polling
miss.\footnote{The weighted-average error of polls within three weeks of
  the 2020 election was 4.2 percentage points in favor of the Democratic
  party candidate, compared to a historical average of 1.3 percentage
  points in favor of the Democratic party candidate (Silver, 2021).}
Some states report most of their results early in the night while others
do not release comprehensive counts until late into the night or even
the following day. This trickle of information provides ample time for
markets to adjust their perceptions of the likeliest outcomes of the
election. In addition, it became obvious that expectations about the
election outcome, which strongly favored Democrats based on polls, had
been inaccurate. The actual election results had a much closer margin
than expected, which forced investors to recalibrate throughout the
night.

Among the important financial markets that clearly react to the events
of the election, Panel B of Figure 1 plots an AFE-USD currency index
(AFE-USD Index) as a percent change from its price prior to 6:00 p.m. on
November 3.\footnote{Our AFE-USD Index measures the exchange rate of the
  dollar against a basket of major advanced foreign economy currencies
  (CAD, CHF, EUR, GBP, JPY, and SEK), borrowing methodology from the
  Federal Reserve Dollar Indexes (von Beschwitz et al., 2019). Likewise,
  our EME-USD Index does the same using emerging market economy
  currencies (CNY, HKD, MXN, SGD, and THB). Currencies were selected for
  inclusion in the indexes based on the availability of minutely price
  data on election night. An increase in the index indicates a
  depreciation of the dollar.} Currency markets are global in nature and
react to major real world events, and unlike equity markets, they trade
overnight, so they are well-positioned to react to information like
election results. The AFE-USD Index clearly reacts to the events of the
November presidential election, with an extremely high correlation of
0.92 between the two series.

\section{Initial Market Reaction}\label{initial-market-reaction}

In Panel A of Figure 2, we plot the Democratic Presidency Probability
against the change in the AFE-USD index across the same time period as
above, 6pm November 3rd through November 4th. The resulting
statistically significant coefficient---highlighted in red---of a simple
linear regression, quantifies the market's pricing of a Democratic over
a Republican victory in the race for the presidency, in this case
implying a depreciation of just over 1\% in the price of the dollar.

Using the same time window of minutely data, we can expand our scope to
a host of markets that are traded overnight: futures for the S\&P 500
index, NASDAQ, and Russell 2000 stock market indexes, the EME-USD Index,
future prices for oil (measured by WTI Crude) and the VIX stock market
volatility index, and the yield of 10-year treasury bonds. Panel B of
Figure 2 plots the resulting coefficients of a linear regression between
each of these financial markets and the Democratic Presidency
Probability, with a 95\% confidence interval shown in bars around each
estimate. Every relationship is statistically significant, and only the
VIX and the 10-Year Treasury yield\footnote{Throughout this note,
  Treasury bond movements are quoted in yields (bond yields rise as
  prices fall.)} have a negative coefficient with the Democratic
Presidency Probability.

\begin{figure}

\centering{

\includegraphics[width=5in,height=\textheight]{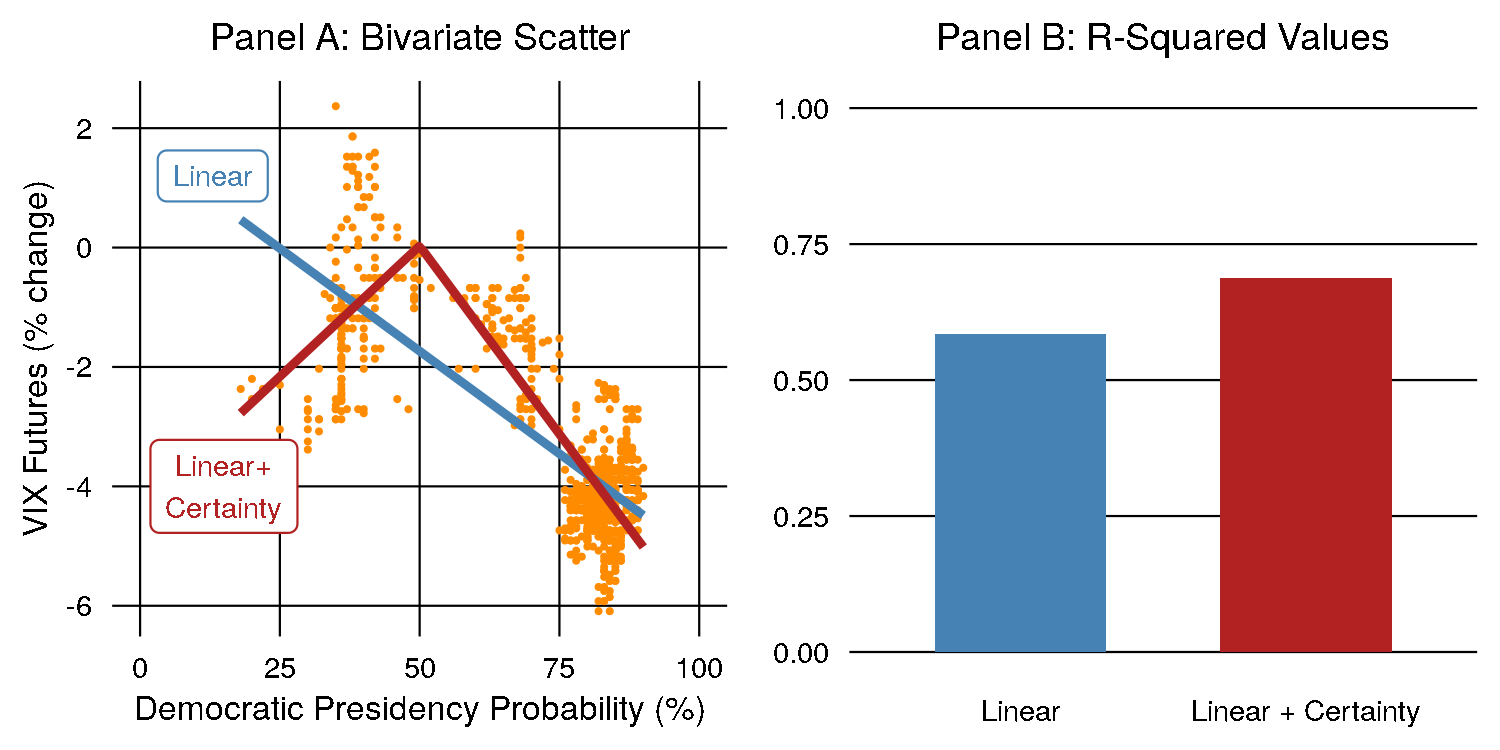}

Note: In Panel A, each point represents one observation of the last
traded price (probability) from PredictIt and the percent change in the
AFE-USD currency index since 6:00 p.m. November 3rd, with higher values
denoting the depreciation of the dollar. The fitted line is a simple
linear regression shown in the upper left. In Panel B, each bar shows
the coefficient between a simple linear regression of one of the
financial markets and the Democratic Presidency Probability, with a 95\%
confidence interval around the estimate.

Source: PredictIt, Bloomberg, authors' calculations.

}

\caption{\label{fig-2}Financial Markets and the Presidency}

\end{figure}%

There are several potential explanations for the observed co-movements
between asset prices and the changing presidency probability. One
possibility is an increasing ``risk-on'' sentiment, where investors are
more willing to own riskier assets in exchange for higher expected
return. Evidence for such a rotation is in the declining VIX futures,
the EME currencies strengthening more than those of AFEs, and the
tech-focused NASDAQ index being the most positively correlated of any of
the assets in our sample. However, a countervailing piece of evidence
against such ``risk-on'' behavior is the negative coefficient in the
10-year treasury yield. A typical ``risk-on'' rotation would involve an
increase in treasury yields as investors sell off these extremely safe
assets. Instead, we observe a negative correlation between treasury
yields and the Democratic presidency probability.

\section{A Reduction in Uncertainty}\label{a-reduction-in-uncertainty}

It is possible that markets were not only pricing in a Democratic or
Republican victory for the presidency, but also the uncertainty of the
election outcome. In face of a large, uncertain event, financial markets
may respond with a ``risk-off'' stance, putting investments into safer
assets until the uncertainty is concluded. After the event has resolved,
markets then take on more risk and lower their expectation of future
volatility.

This pricing-in of uncertainty is clear in the VIX futures price,
plotted against the Democratic Presidency Probability in Panel A of
Figure 3. The simple linear regression, in blue, has a fairly strong fit
to the data, as demonstrated by its R-squared value of about 0.6,
plotted in Panel B. However, by adding an additional variable capturing
the certainty of the election outcome (calculated as the absolute value
of the distance from 50-50 odds), the model fit of a specification we
label ``Linear + Certainty'' is even higher. The strength of the
relationship is also visually convincing in Panel A of Figure 3, with
the ``Linear + Certainty'' model plotted with the red line. This
suggests that to some extent, market participants were responding to the
level of certainty of the presidential outcome instead of the outcome
itself.

\begin{figure}

\centering{

\includegraphics[width=5in,height=\textheight]{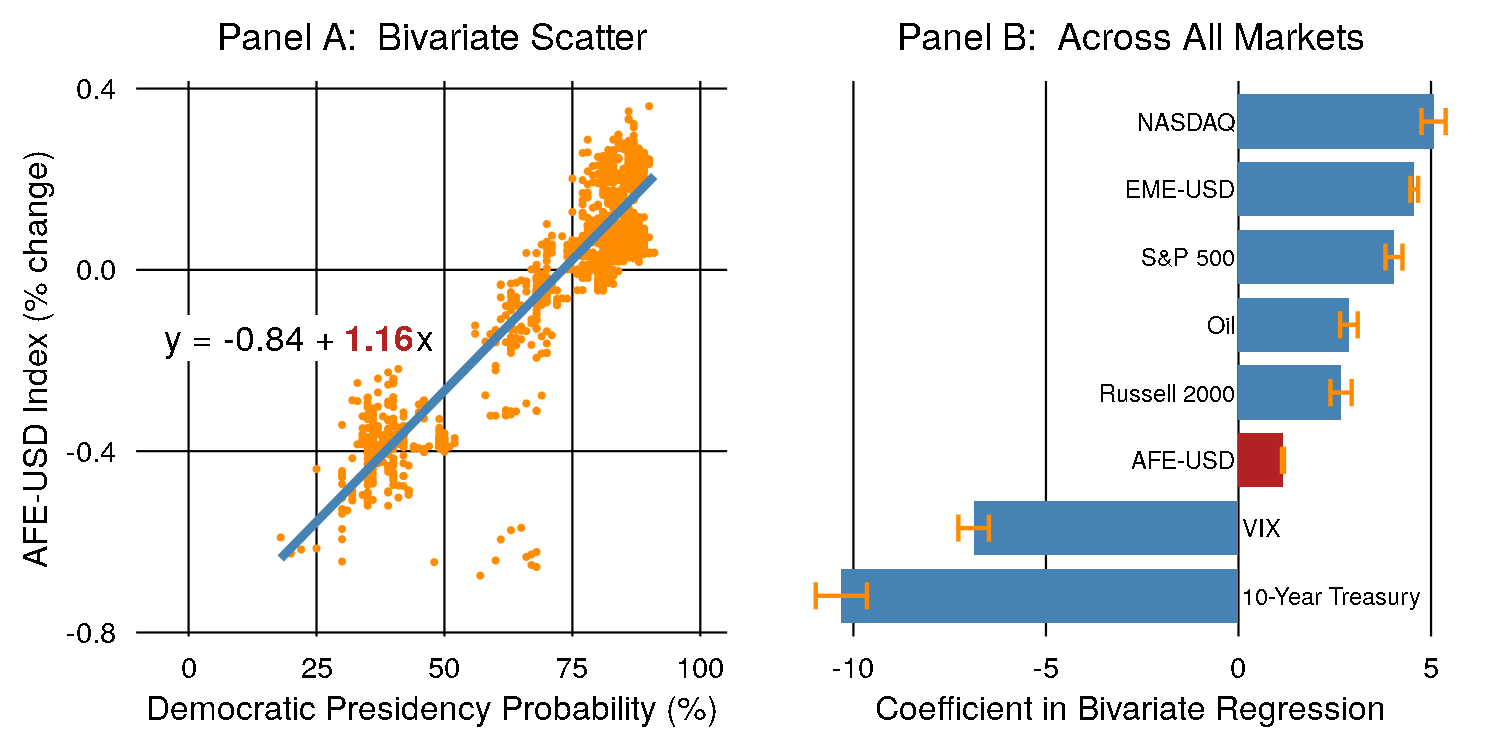}

Note: In Panel A, each point represents one observation of the last
traded price (probability) from PredictIt and the percent change in
price of the VIX futures since 6:00 p.m. November 3rd. The fitted line
in blue is a simple linear regression and the line in red has an
additional term, the absolute value of the distance from 50-50 odds. In
Panel B, each bar shows the R-Squared of the two models fitted in Panel
A.

Source: PredictIt, Bloomberg, authors' calculations.

}

\caption{\label{fig-3}Adding Certainty to the Equation}

\end{figure}%

\section{Controlling for Other
Elections}\label{controlling-for-other-elections}

Earlier we noted that by narrowing the time window to election night, we
were reducing the probability that other events could be affecting
financial markets. Our analysis so far, however, abstracts from a host
of other elections that take place at the same time as the U.S.
presidential election. There are two other crucial outcomes: control of
the Senate and of the House of Representatives. Democrats were the
favorites to win control of the House, and the probability of that
outcome did not move much over the course of the night (ranging from
88\% to 99\%). Control of the Senate, on the other hand, moved
substantially, favoring Democrats heading into election night (about
70\%) and completely switching by the end of the following day with a
Republican win probable (about 80\%). An additional market on PredictIt
asked what the probability was of a Democratic ``Trifecta'' across the
federal government, with control of the Senate, House, and Presidency.
The resulting coefficients of regressing various financial markets on
these four variables---Democratic Presidency Probability, Presidential
Certainty, Democratic Senate Probability, and Democratic Trifecta
Probability---are shown in Figure 4, which paints the most complete
picture of how the various elections of 2020 were priced in by
markets.\footnote{For the purposes of our analysis, the inclusion of the
  Trifecta is similar to adding an interaction term between the
  Democratic Presidency and Democratic Senate probabilities.}

\begin{figure}

\centering{

\includegraphics[width=5in,height=\textheight]{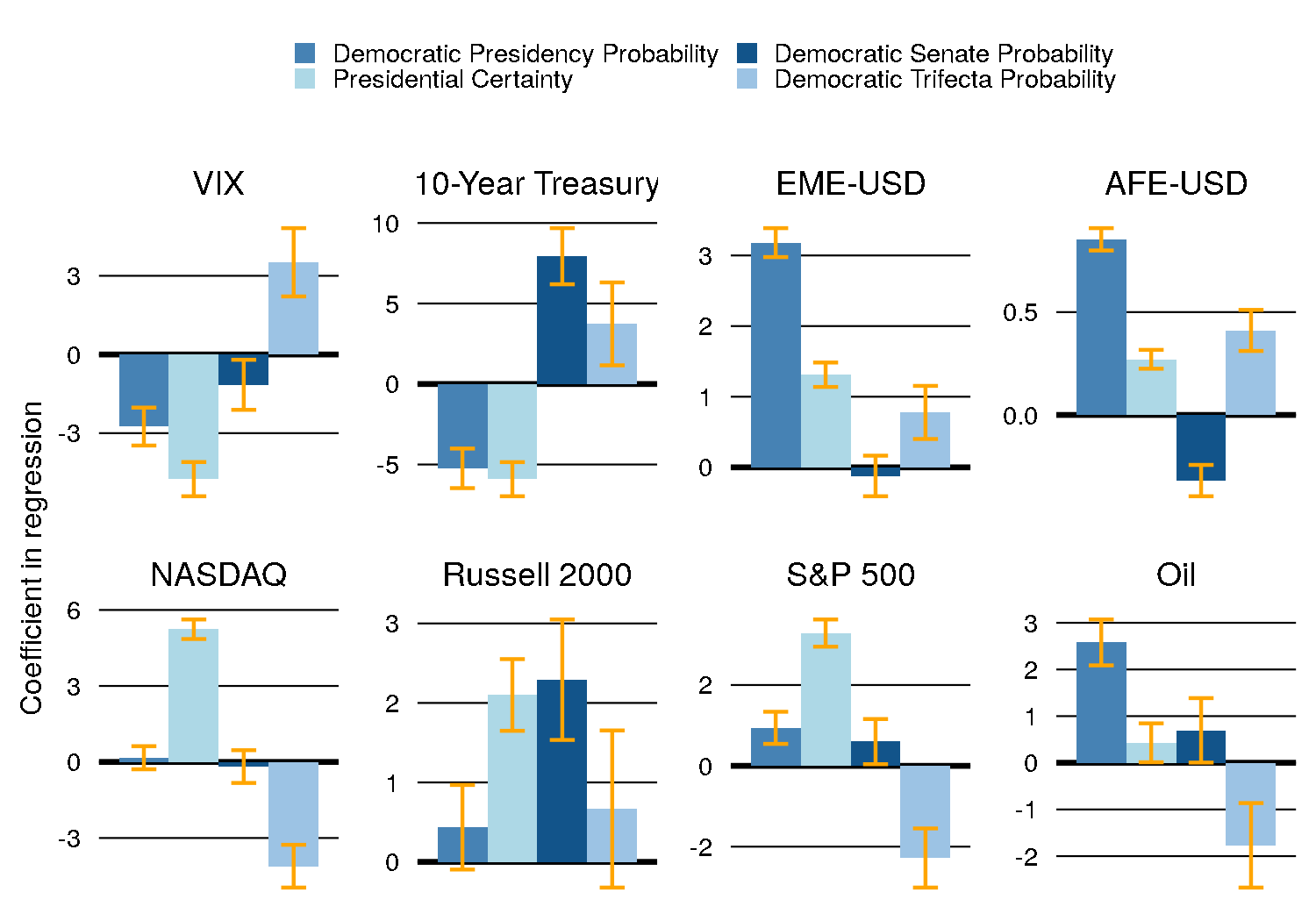}

Note: Each panel shows four bars representing the coefficients on a
multivariable regression between the four prediction market variables
and each of the financial variables. The yellow error bars show the 95\%
confidence interval around each estimate.

Source: PredictIt, Bloomberg, authors' calculations.

}

\caption{\label{fig-4}A Multivariable Picture of Markets' Reactions}

\end{figure}%

Looking first at the VIX, we can see that the coefficient on
Presidential Certainty captures the largest amount of movement in the
financial market dependent variables, with the Democratic Presidency
Probability still having a large negative effect, and the Democratic
Trifecta Probability associated with an increase in the VIX. It is
likely that the shock effect of one party controlling all three
institutions, which was a surprise outcome, led to expectations of
responsive volatility in equity markets in the near term. In contrast,
the 10-Year Treasury shows a decrease in yield from both of the
Presidential variables (as is consistent with the results in Figure 2),
but it has an implied increase in yield from the Democratic Senate and
Trifecta Probabilities. This is a possible sign that markets interpreted
a Democratic Senate (and even more so a Trifecta) as a precursor of a
larger round of stimulus owing to a unified government being more easily
able to pass spending legislation.

In currency markets, the AFE-USD currency index shows the relatively
lowest absolute movement in response to the election, while the EME-USD
currency index has one of the strongest absolute responses. As before,
there is a much larger implied appreciation of EME currencies compared
to AFE currencies for an increase in both Democratic Presidential
Probability and Presidential Certainty, pointing to ``risk-on''
behavior. Both currency indexes have a positive and significant
coefficient on Democratic Trifecta Probability, and the AFE index has a
significant negative coefficient on Democratic Senate Probability.

The stock market index futures exhibit different behaviors, with the
Russell 2000 increasing in response to each of the variables, but
especially to the probability of a Democratic Senate as well as
Presidential Certainty. The NASDAQ, on the other hand, has a large
implied increase on Presidential Certainty, but is negatively correlated
with Democratic Trifecta Probability. The S\&P 500 shows similar
reactions to the NASDAQ but also has significant and positive (though
small) coefficients on Democratic Presidency Probability and Democratic
Senate Probability.

Oil responded most positively to the Democratic Presidency Probability
and has one of the smallest coefficients with Presidential Certainty,
suggesting that the market is interpreting some potential policy
differences between the two presidential outcomes.

\section{The Georgia Runoffs and Control of the
Senate}\label{the-georgia-runoffs-and-control-of-the-senate}

The November election left the Senate split with 50 Republican seats to
48 for the Democrats.\footnote{This includes the two independents who
  caucus with the Democrats in the Democratic total.} The final two
seats were both in Georgia as no candidate won 50\% of the vote,
triggering automatic runoffs that would decide control of the Senate,
with Democrats only needing 50 seats to hold a majority (using the Vice
President's vote to break ties).

We repeat the analysis above, focusing instead on January 5th from 6:00
p.m. to midnight, as the final results of the election came in. This
time there truly was only one election outcome (two races) and the
result was unexpected, causing a steady shift in the odds as the votes
were tallied. In Panel A, the strong relationship between the Democratic
Senate Probability and the 10-Year Treasury Price is apparent. The
strong increase in the VIX paints a picture of the surprise outcome
introducing volatility into the market as prices adjust. In addition,
the movements may reveal an expectation of a larger stimulus package
because of the prospect of a unified government being able to pass bills
more easily (consistent with results in Figure 4). Most notably, the
implied increase in the 10-Year Treasury is quite significant and aligns
well with this story, on the textbook macroeconomic assumption that
higher government spending leads to higher future interest rates. The
opposite reactions of the Russell 2000 compared to the S\&P 500 and
NASDAQ are also revealing, again showing the importance of this
political event to the financial markets.

\begin{figure}

\centering{

\includegraphics[width=5in,height=\textheight]{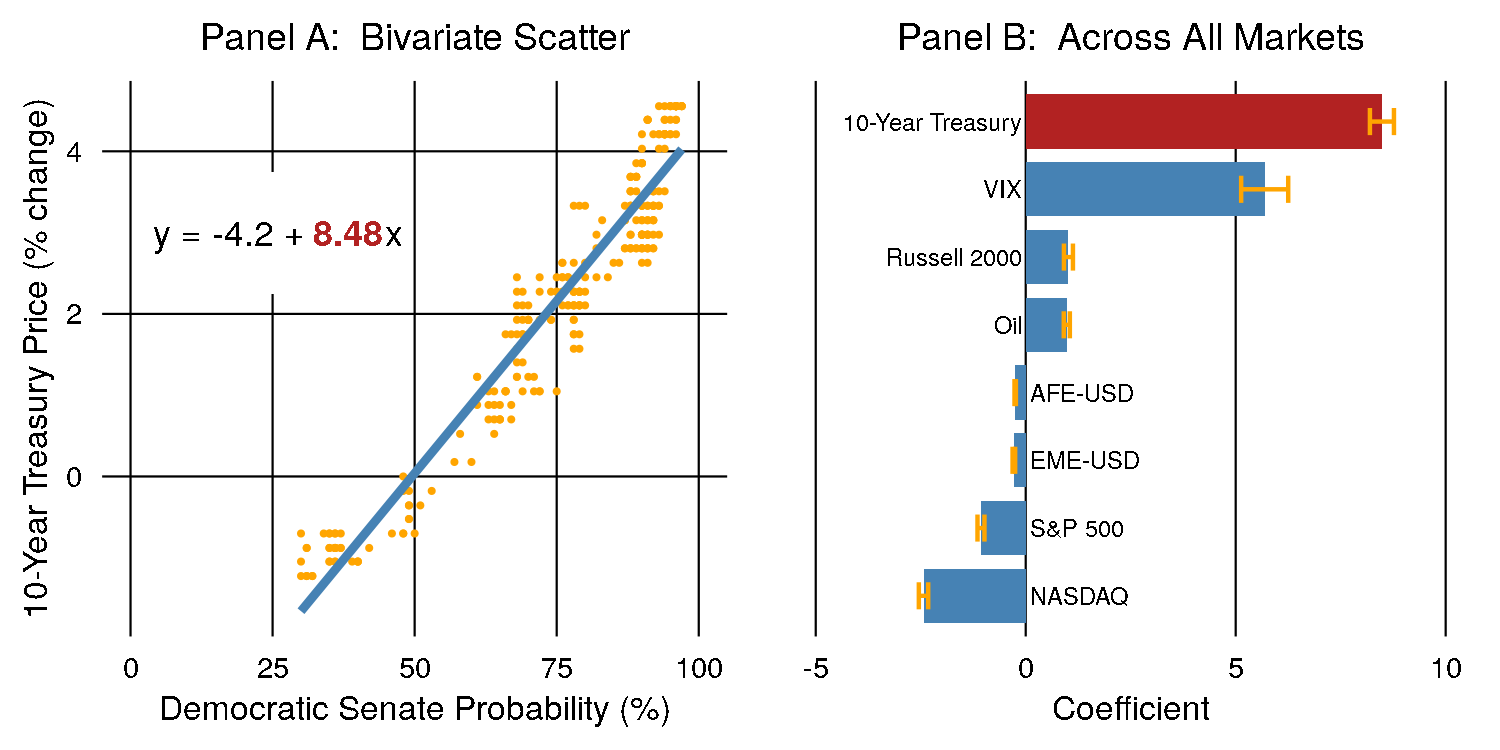}

Note: In Panel A, each point represents one observation of the last
traded price (probability) from PredictIt and the percent change in
price of the 10-Year Treasury since 6:00 p.m. January 5th. The fitted
line in blue is a simple linear regression shown in the upper left. In
Panel B, each bar shows the coefficient between a simple linear
regression of one of the financial markets and the Democratic Senate
Probability, with a 95\% confidence interval around the estimate.

Source: PredictIt, Bloomberg, authors' calculations.

}

\caption{\label{fig-5}Financial Markets and Control of the Senate}

\end{figure}%

\section{Conclusion}\label{conclusion}

We use a novel minutely data set to provide insight into the precise
relationship between financial markets and the changes in 2020 election
result probabilities. While we cannot identify the exact reasons for any
of the market movements throughout the course of the election, the
striking correlations show that prediction markets and financial markets
are indeed tightly interlinked. This result shows the importance of
further studying the utility of prediction markets to make economic and
political forecasts.

\section{References}\label{references}

Arrow, K. J., Forsythe, R., Gorham, M., Hahn, R., Hanson, R., Ledyard,
J. O., Levmore, S., Litan, R., Milgrom, P., Nelson, F. D., Neumann, G.
R., Ottaviani, M., Schelling, T. C., Shiller, R. J., Smith, V. L.,
Snowberg, E., Sunstein, C. R., Tetlock, P. C., Tetlock, P. E., \ldots{}
Zitzewitz, E. (2008). The Promise of Prediction Markets. Science,
320(5878), 877--878. https://doi.org/10.1126/science.1157679

Bloomberg Finance LP. Bloomberg Terminals (Open, Anywhere, and Disaster
Recovery Licenses).

Hahn, R. W., \& Tetlock, P. C. (2005). Using Information Markets to
Improve Public Decision Making. Harvard Journal of Law \& Public Policy,
29, 213.

Hanson, R. (1996, June 9). Idea Futures---The Concept.
https://mason.gmu.edu/\textasciitilde rhanson/ideafutures.html

O'Leary, D. E. (2015). User participation in a corporate prediction
market. Decision Support Systems, 78, 28--38.
https://doi.org/10.1016/j.dss.2015.07.004

Plott, C. R., \& Chen, K.-Y. (2002, March). Information Aggregation
Mechanisms: Concept, Design and Implementation for a Sales Forecasting
Problem (Report or Paper No.~1131). California Institute of Technology.
https://resolver.caltech.edu/CaltechAUTHORS:20140317-135547085

PredictIt, Market Data via API,
https://www.predictit.org/api/marketdata/all/

Reade, J. J., \& Vaughan Williams, L. (2019). Polls to probabilities:
Comparing prediction markets and opinion polls. International Journal of
Forecasting, 35(1), 336--350.
https://doi.org/10.1016/j.ijforecast.2018.04.001

Rhode, P. W., \& Strumpf, K. S. (2004). Historical Presidential Betting
Markets. Journal of Economic Perspectives, 18(2), 127--141.
https://doi.org/10.1257/0895330041371277

Silver, N. (2021, March 25). The Death Of Polling Is Greatly
Exaggerated. FiveThirtyEight.
https://fivethirtyeight.com/features/the-death-of-polling-is-greatly-exaggerated/

Snowberg, E., Wolfers, J., \& Zitzewitz, E. (2013). Chapter
11---Prediction Markets for Economic Forecasting. In G. Elliott \& A.
Timmermann (Eds.), Handbook of Economic Forecasting (Vol. 2,
pp.~657--687). Elsevier.
https://doi.org/10.1016/B978-0-444-53683-9.00011-6

Spears, B., LaComb, C., Interrante, J., Barnett, J., \&
Senturk-Dogonaksoy, D. (2009). Examining Trader Behavior in Idea
Markets: An Implementation of GE's Imagination Markets. Journal of
Prediction Markets, 3(1), 17--39.

von Beschwitz, Bastian, Christopher G. Collins, and Deepa D. Datta
(2019). ``Revisions to the Federal Reserve Dollar Indexes,'' FEDS Notes.
Washington: Board of Governors of the Federal Reserve System, January
2019., https://doi.org/10.17016/2573-2129.48.

Wolfers, J. (2009). Prediction Markets: The Collective Knowledge of
Market Participants. Cfa Institute Conference Proceedings Quarterly, 26,
37--44. https://doi.org/10.2469/cp.v26.n2.2

\end{document}